# Optimizing Text Search: A Novel Pattern Matching Algorithm Based on Ukkonen's Approach

Xinyu Guan BAIDU xinyuguanphd@outlook.com
Shaohua Zhang The university of Glasgow zhangshtj2022@163.com

**Abstract—In the realm of computer science, the efficiency of text-search algorithms is crucial for processing vast amounts of data in areas such as natural language processing and bioinformatics. Traditional methods like Naive Search, KMP, and Boyer-Moore, while foundational, often fall short in handling the complexities and scale of modern datasets, such as the Reuters corpus and human genomic sequences. This study rigorously investigates text-search algorithms, focusing on optimizing Suffix Trees through methods like Splitting and Ukkonen's Algorithm, analyzed on datasets including the Reuters corpus and human genomes. A novel optimization combining Ukkonen's Algorithm with a new search technique is introduced, showing linear time and space efficiencies, outperforming traditional methods like Naive Search, KMP, and Boyer-Moore. Empirical tests confirm the theoretical advantages, highlighting the optimized Suffix Tree's effectiveness in tasks like pattern recognition in genomic sequences, achieving 100% accuracy. This research not only advances academic knowledge in text-search algorithms but also demonstrates significant practical utility in fields like natural language processing and bioinformatics, due to its superior resource efficiency and reliability.**

## I. INTRODUCTION

In an era marked by the digital information explosion, the vast amounts of unstructured text data generated across various sectors necessitate advanced algorithmic approaches for efficient retrieval and analysis [1]. This trend is particularly pronounced in domains like natural language processing, where the subtleties of human language add layers of complexity, and in bioinformatics, where the sheer volume of genomic data demands robust search capabilities. Data analytics, another field greatly impacted, relies heavily on extracting meaningful insights from unstructured data. The overarching challenge in these domains is to develop algorithms that not only process data efficiently but also maintain high accuracy in pattern recognition, a balance that is crucial yet elusive in the computational world.

In addressing these computational challenges, our study zeroes in on the utilization of Suffix Trees, acclaimed for their efficiency in text search and analysis. The construction of these trees in our research is uniquely anchored in Ukkonen's Algorithm, renowned for its linear time efficiency in building Suffix Trees [3]. This choice reflects our commitment to optimizing the foundational aspects of our data structure. Building upon this, the research introduces a groundbreaking innovation: a novel search algorithm specifically tailored for use on Suffix Trees. This newly designed algorithm is a departure from conventional methods, offering rapid and efficient searching capabilities.

In the specific design process of our novel algorithm, we have pioneered the use of a unique treenode data structure, leveraging Python's dynamic capabilities, particularly its link attributes [2]. This innovative approach plays a crucial role in the construction of the Suffix Trees. During the building phase, we ensure that key information is interlinked, establishing connections between various important elements of the tree. This interconnected structure of treenodes facilitates the accessibility and integration of critical data, such as node relationships and tree depth, which are essential for efficient searching.

The algorithm ’ s design process also incorporates dynamic computation of leaf nodes and coordinate parameters, pivotal for enhancing search speed. By calculating these elements on-the-fly, our algorithm can rapidly navigate through the Suffix Tree, identifying relevant patterns with increased precision and speed. This method contrasts traditional search algorithms that may not utilize such dynamic calculations, often resulting in slower search times and increased complexity, especially when dealing with large datasets.

Furthermore, the incorporation of Python's link attributes in our treenode design allows for a more fluid and adaptable algorithm. This adaptability is particularly beneficial when handling diverse and complex data structures, as it enables the algorithm to efficiently manage and traverse the hierarchical nature of Suffix Trees. As a result, our innovative search algorithm not only capitalizes on the inherent strengths of Ukkonen's Algorithm but also introduces a new level of efficiency in text-search operations, marking a significant advancement in the field [4].

The pivotal innovation of this research lies in an advanced algorithmic optimization that synergizes the strengths of Ukkonen's Algorithm with a custom-designed search algorithm. This hybrid approach is meticulously crafted to transcend the existing limits of computational efficiency in text-search algorithms. The optimization is expected to reduce the time complexity and improve the space efficiency of Suffix Tree operations, thereby setting a new benchmark in the field of text-search algorithms.

To contextualize the advancements of this study, a comprehensive comparative analysis is conducted, encompassing the Python implementations of four widely-recognized text-search algorithms: Naive String Search, KMP (Knuth-Morris-Pratt) String Search, Rabin-Karp String Search, and BM (Boyer-Moore) String Search. This analysis is not merely theoretical; it extends to empirical evaluations that critically examine the time and space complexities of these algorithms under diverse real-world scenarios. This comparative study is essential to gauge the relative performance and practical applicability of each algorithm, thereby validating the superiority of the proposed method.

The comprehensive complexity analyses conducted in this study reveal a notable finding: Ukkonen's Algorithm exhibits linear time and space complexities [3]. This is a significant advancement over traditional algorithms like Naive Search, KMP, and Boyer-Moore, which often have higher complexities. To reinforce these theoretical findings, empirical evaluations were systematically carried out using diverse datasets, such as the Reuters corpus for textual data and various human genomic sequences. These evaluations not only corroborate the theoretical predictions but also demonstrate the practical efficacy of the algorithm in handling large-scale data efficiently.

## II. RELATED WORK

String matching, essential in text processing, plays a pivotal role in diverse applications such as operating systems, bioinformatics, information retrieval, and cybersecurity [5]. The development of efficient string matching algorithms, both single-pattern and multi-pattern, is crucial for managing large-scale data and facilitating complex queries.

In the realm of single-pattern matching algorithms, essential for locating unique patterns within large texts, the journey began with the Knuth-Morris-Pratt (KMP) algorithm in the 1970s [6]. This algorithm introduced a preprocessing step to enhance search efficiency. Later, the Boyer-Moore (BM) algorithm [7] further optimized searches for smaller character sets but encountered challenges with larger sets like Unicode. This led to enhancements by ZR Feng and T Takaoka [8] and applications in areas such as web application vulnerability detection by Ain Zubaidah Mohd Saleh's team [9]. Concurrently, the Rabin-Karp algorithm [10] used hashes for efficient matching in large data volumes, although it sometimes produced false positives.

On the other hand, multi-pattern matching algorithms, which handle multiple patterns simultaneously, saw significant advancements with the Aho-Corasick algorithm in 1975 [11]. This algorithm integrated Finite State Machines for efficient synchronous pattern matching. The Commentz-Walter method [12] later combined Boyer-Moore's skipping technique with Aho-Corasick's automaton structure, enhancing adaptability but facing high time complexity in some scenarios. The Baeza-Yates approach [13] merged Boyer-Moore-Horspool with Aho-Corasick, proving effective for large datasets and real-time applications.

The field's progression led to the integration of suffix structures such as Tries and Trees in string matching. A notable development was Ukkonen's linear-time suffix tree construction algorithm [14], which significantly reduced space complexity. G Navarro's Splitting Method [15] optimized this approach by segmenting text or patterns for more efficient matching.

After reviewing both single-pattern and multi-pattern matching algorithms, this study identifies the challenges in processing complex texts and large datasets. Addressing these, we propose a novel string searching algorithm based on Ukkonen's suffix tree construction. This method aims to enhance spatial and temporal efficiency and precision in processing large-scale text data.

## III. METHODOLOGY

This section of the thesis delineates the comprehensive methodologies adopted for refining text search algorithms. The study is bifurcated into two distinct but interrelated parts: the first part comprises an evaluative comparison of established single-pattern search algorithms, serving as a baseline for experimental benchmarking. The second part, forming the nucleus of the research, delves into the advanced realms of multi-pattern algorithms, with a particular emphasis on the Suffix Tries and the enhanced Suffix Tree Search Algorithms.

### A. Evaluative Comparison of Single-pattern Search Algorithms.

In the realm of single-pattern search algorithms, several methodologies have been scrutinized to establish a baseline for comparison. These include the Naive String Search, Rabin-Karp, Knuth-Morris-Pratt (KMP), and Boyer-Moore (BM) algorithms. Each of these algorithms presents a unique interplay of computational efficiency and algorithmic complexity.

The Naive String Search, known for its straightforward implementation, stands out for its simplicity but falls short in terms of efficiency, particularly when dealing with extensive texts. In contrast, the Rabin-Karp algorithm, employing hash functions, offers a significant improvement in speed, especially in scenarios involving multiple pattern searches. However, its susceptibility to hash collisions and relatively slower performance in single pattern searches cannot be overlooked.

As illustrated in Fig. 1, the KMP algorithm, with its ingenious utilization of the Longest Prefix Suffix (LPS) array, effectively eliminates unnecessary comparisons, thereby optimizing the search process. Nevertheless, the complexity in understanding and implementing the LPS array is a notable limitation. As its data structure is shown in Figure 2 , the Boyer-Moore algorithm, arguably the most efficient in this category, especially for longer patterns, leverages heuristic-based techniques to skip sections of the text, significantly reducing the number of comparisons. However, its complexity in preprocessing and diminished effectiveness for short patterns and limited alphabet sets present certain constraints.

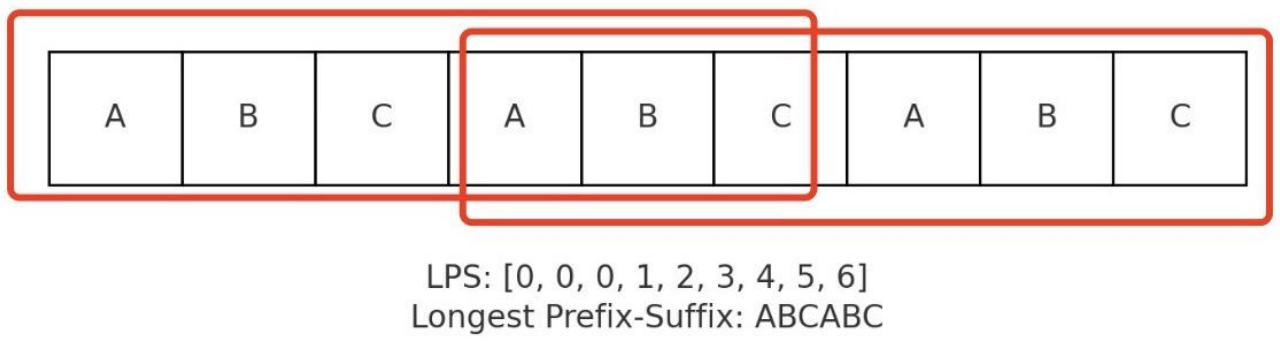

Fig. 1. KMP algorithm data structure

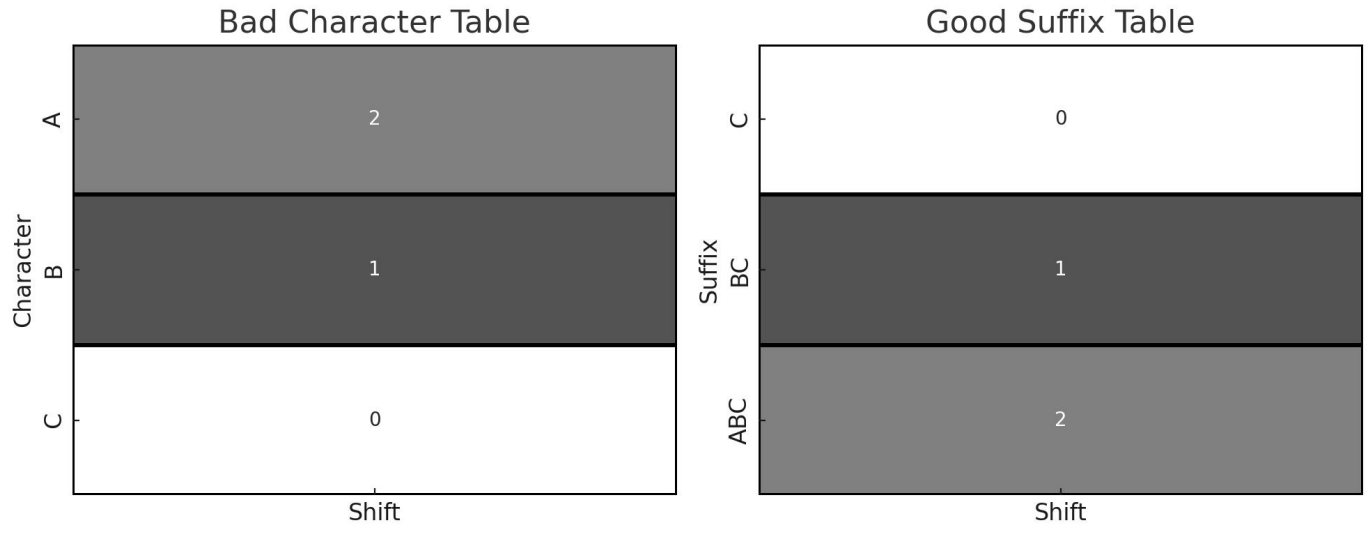

Fig. 2. BM algorithm data structure

### B. Core Research: Multi-pattern Search Algorithms

This segment of the thesis extensively discusses the development and nuances of multi-pattern search algorithms, with a special focus on the Suffix Trie and Suffix Tree algorithms, as well as the introduction of a novel framework proposed in this study.

The Suffix Trie and Suffix Tree algorithms are pivotal in the domain of text search, primarily due to their efficient tree construction mechanisms. Both algorithms construct tree structures that represent various suffixes of a given text, facilitating rapid and efficient search operations. The Suffix

Trie is renowned for its straightforward approach to storing every suffix of the text as a path from the root of the tree, making it highly accessible but space-intensive.Its tree structure is shown in Figure 4. On the other hand, the process of suffix tree generation as shown in Figure 3, an optimized version of the Suffix Trie, compresses these paths into a more compact form, significantly reducing space requirements while maintaining quick search capabilities.

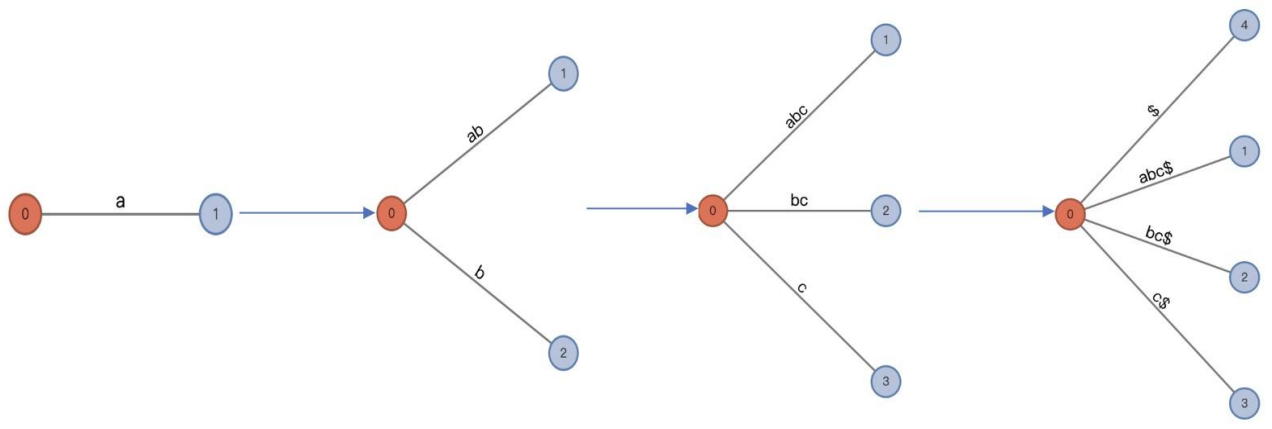

Fig. 3. Suffix tree generation

Building on these foundational concepts, the thesis introduces a novel framework designed to enhance the search functionality on Suffix Trees, particularly in Python. The proposed framework is structured into two main stages,The flowchart is shown in Figure 5:

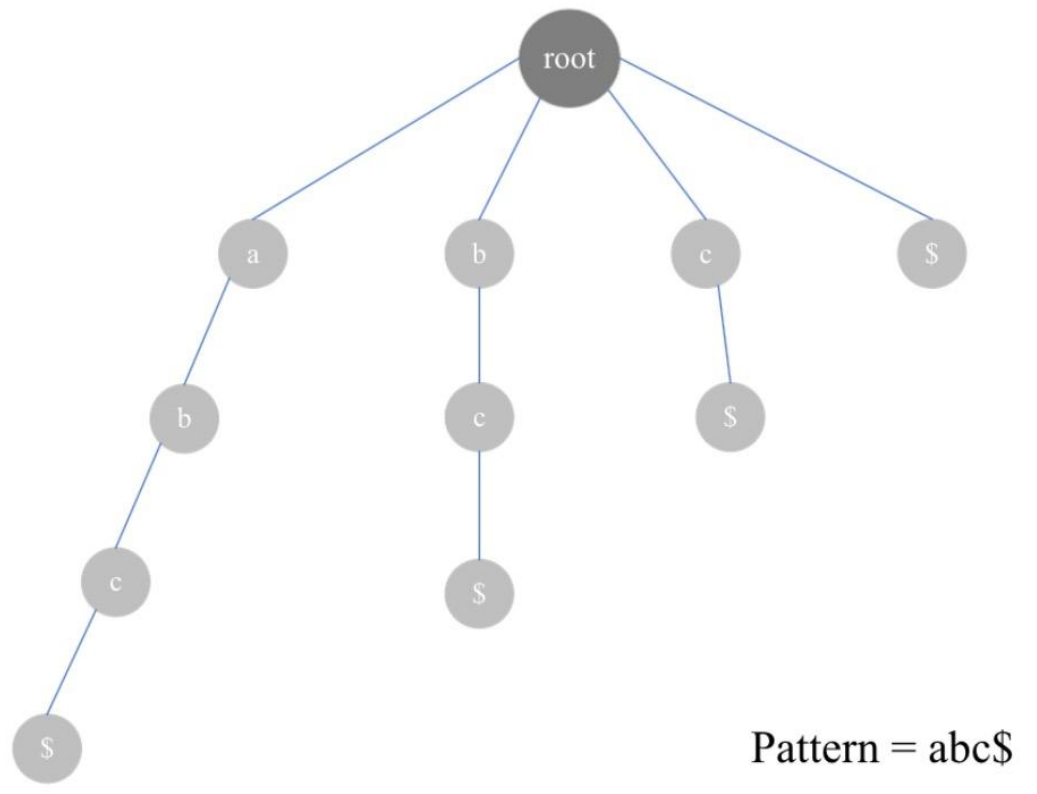

Fig. 4. Suffix tries structure

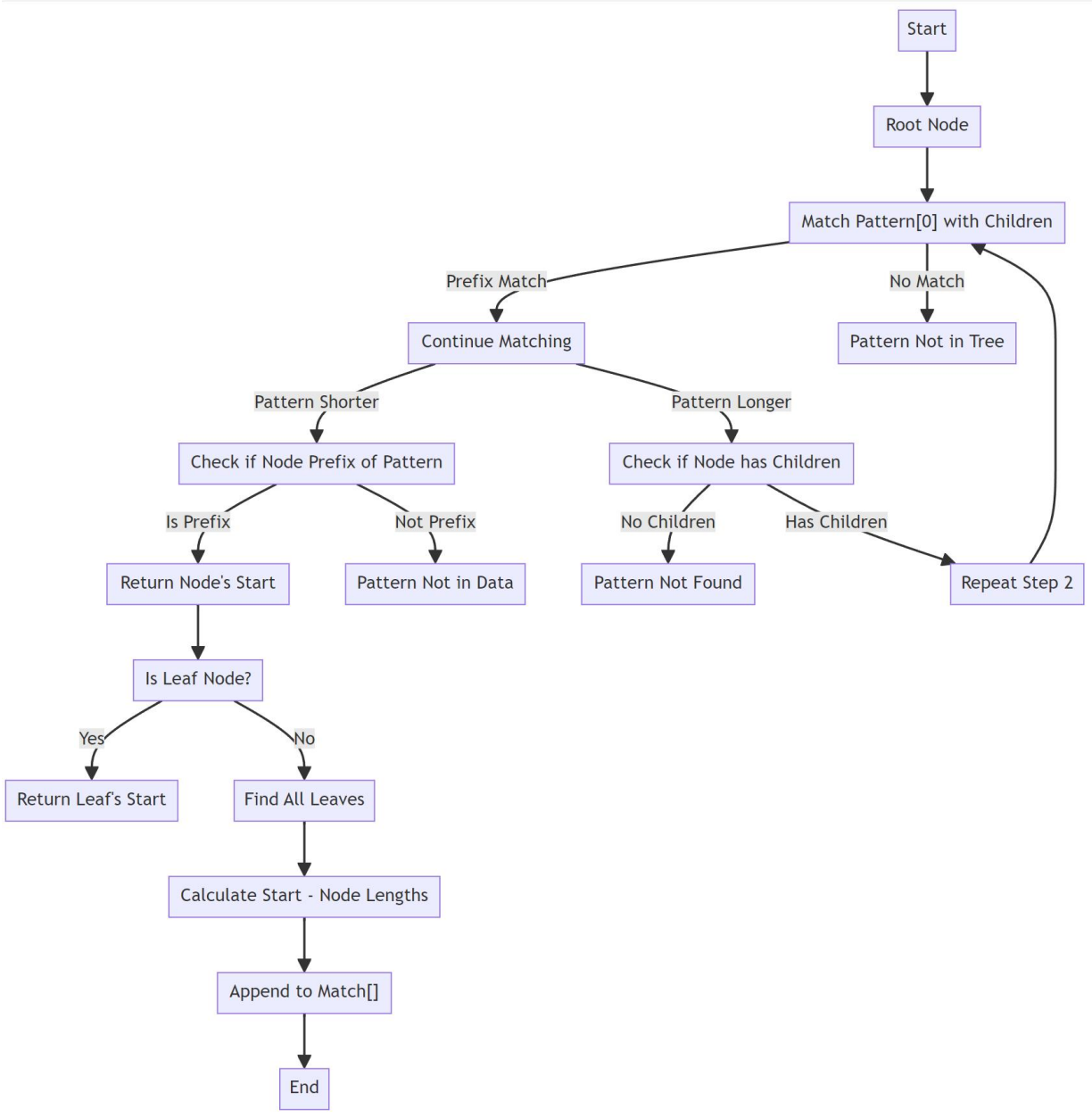

Fig. 5. Search algorithm flow chart

1) *Pattern Matching Node Identification:*

The algorithm begins at the root, matching the pattern with node prefixes and progressing through children as necessary. Upon finding the matching node, it computes the start indices for the pattern within the text. This process, efficient and streamlined, ensures precise pattern location and identification. (Refer to Fig. 5 for visual representation.)

2) *Traversal and Index Calculation:*

This stage involves traversing the subtree from the matching node. If it's a leaf, the start index is directly used. For non-leaf nodes, the algorithm calculates indices of all underlying leaves, adjusting them based on cumulative node depths, thus pinpointing the pattern's exact start positions in the text.

The innovative design of this framework lies in its integrated approach to pattern matching and index calculation within Suffix Trees, enhancing efficiency and accuracy. This Python-based solution is particularly effective for large datasets, offering precise pattern location and adaptability to both leaf and non-leaf nodes.

## IV. EXPERIMENT

### *A. Experimental Settings*

1) *Datasets:*

- Natural Language Texts: The Reuters corpus from NLTK, featuring several thousand articles and millions of tokens, was used to evaluate text search algorithms in NLP applications. A word cloud in Figure 6 highlights the most frequent terms.

- Genomic Sequences: Human genomic sequences were included to present computational challenges. These sequences, rich in unique patterns, were crucial for testing algorithmic performance in bioinformatics. Figures 7 illustrate the frequency distribution of the bases.

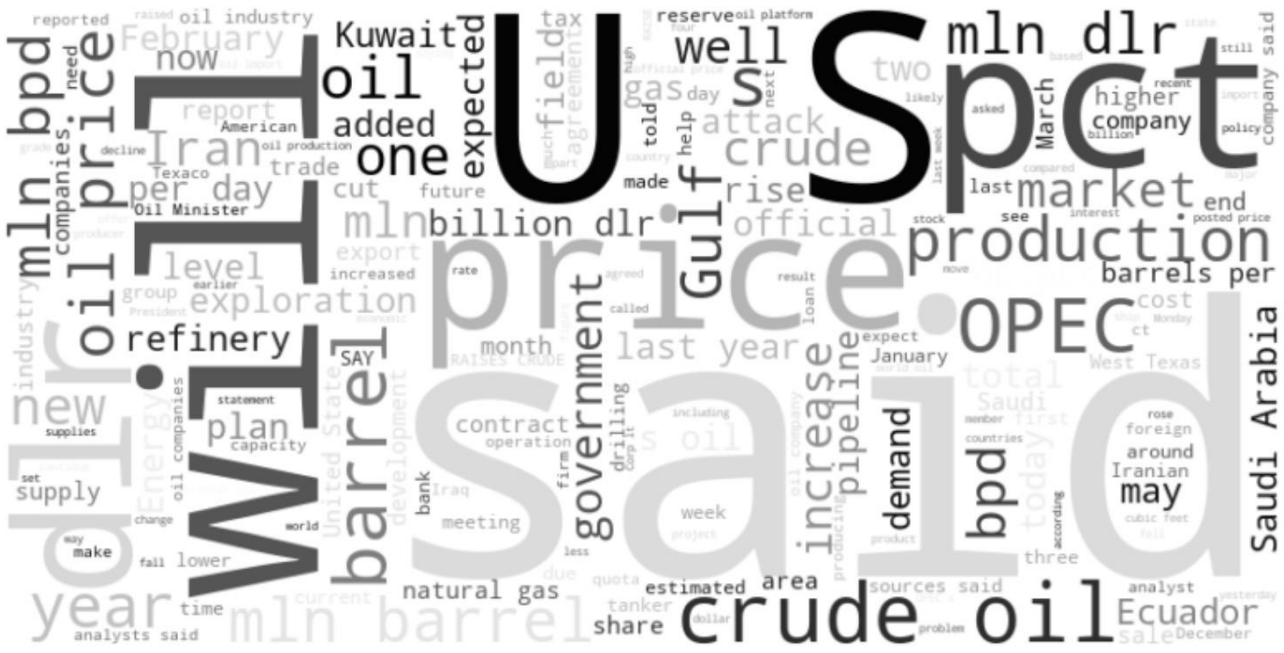

Fig. 6. The most frequent terms

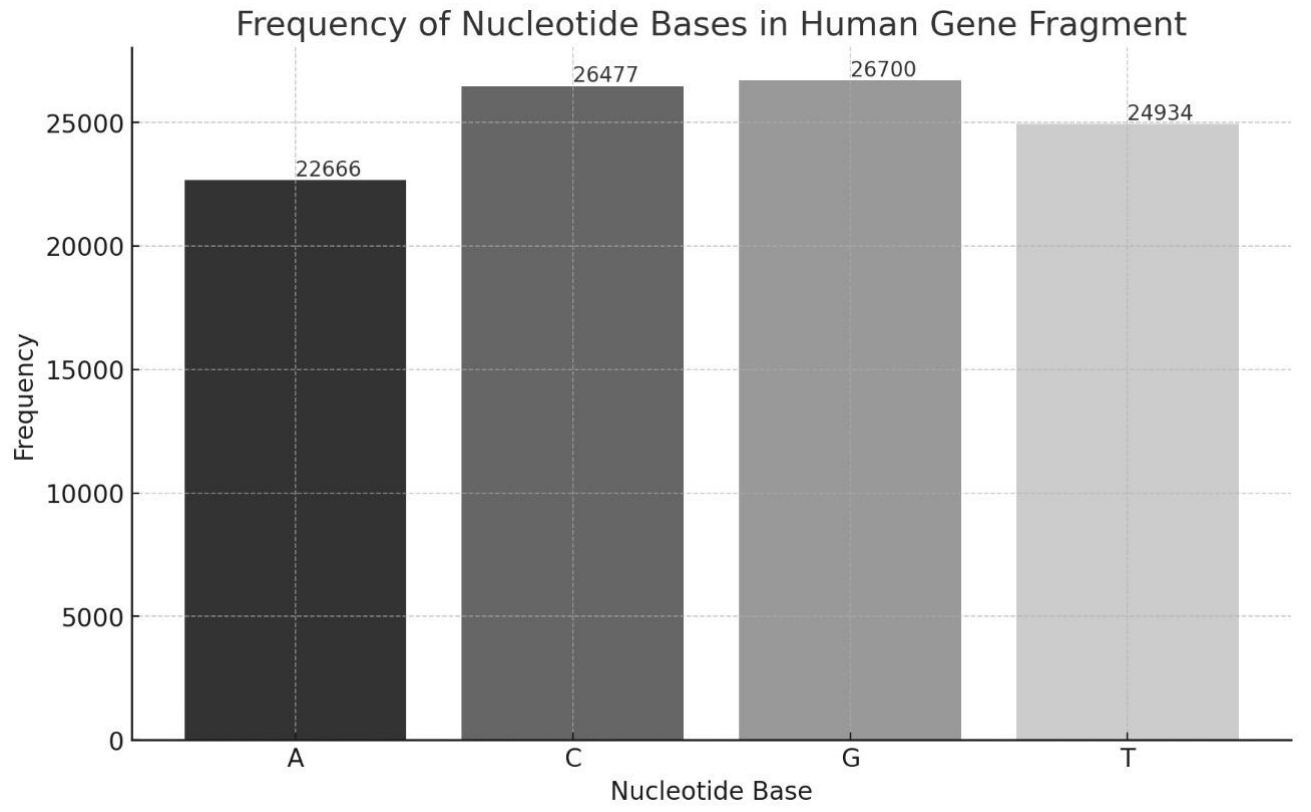

Fig. 7. frequency distribution of the bases.

*2) Metrics:*

In the study, the evaluation of algorithmic efficiency was conducted by measuring the execution time across various pattern lengths and dataset sizes. System resource utilization was assessed with a focus on tree node count, CPU utilization, and memory usage. For accuracy analysis, the performance of multiple text-search algorithms was compared against a gold standard to ensure reliability and precision in results.

*3) Baseline:*

In this experiment, three classical baseline algorithms of kmp, rabin-karp and BM were used to compare our newly designed suffix tree search methods.

### *B. Implement Details & main results:*

*1) Speed test:*

In the algorithmic efficiency assessment, various observations emerged from the analysis of Figures 8, 9, 10, and 11. Initially, Figure 8 indicated that the Suffix Tree Search method was slower in scenarios with lower character counts, likely due to the time required to traverse the nodes of the suffix tree. However, Figures 9 and 10 revealed that in datasets ranging from 250 to 500 characters, the Suffix Tree Search performed comparably to other algorithms. Notably, within this range, the Rabin-Karp algorithm exhibited a rapid increase in computational time. As the dataset sizes expanded beyond 500 characters, the newly optimized Suffix Tree algorithm demonstrated its strengths, generally demanding the least computational time, thus proving effective for larger datasets. A key observation from Figure 11, which tested a dataset of 10,000 characters, was that the improved Suffix Tree algorithm significantly outperformed traditional algorithms like KMP, BM, and Rabin-Karp by over 100 times. Intriguingly, the Rabin-Karp algorithm was the slowest, a finding that contradicts theoretical expectations and might be attributed to variations in Python's hash() function performance across different CPU architectures, such as Apple's ARM and Intel's x86.

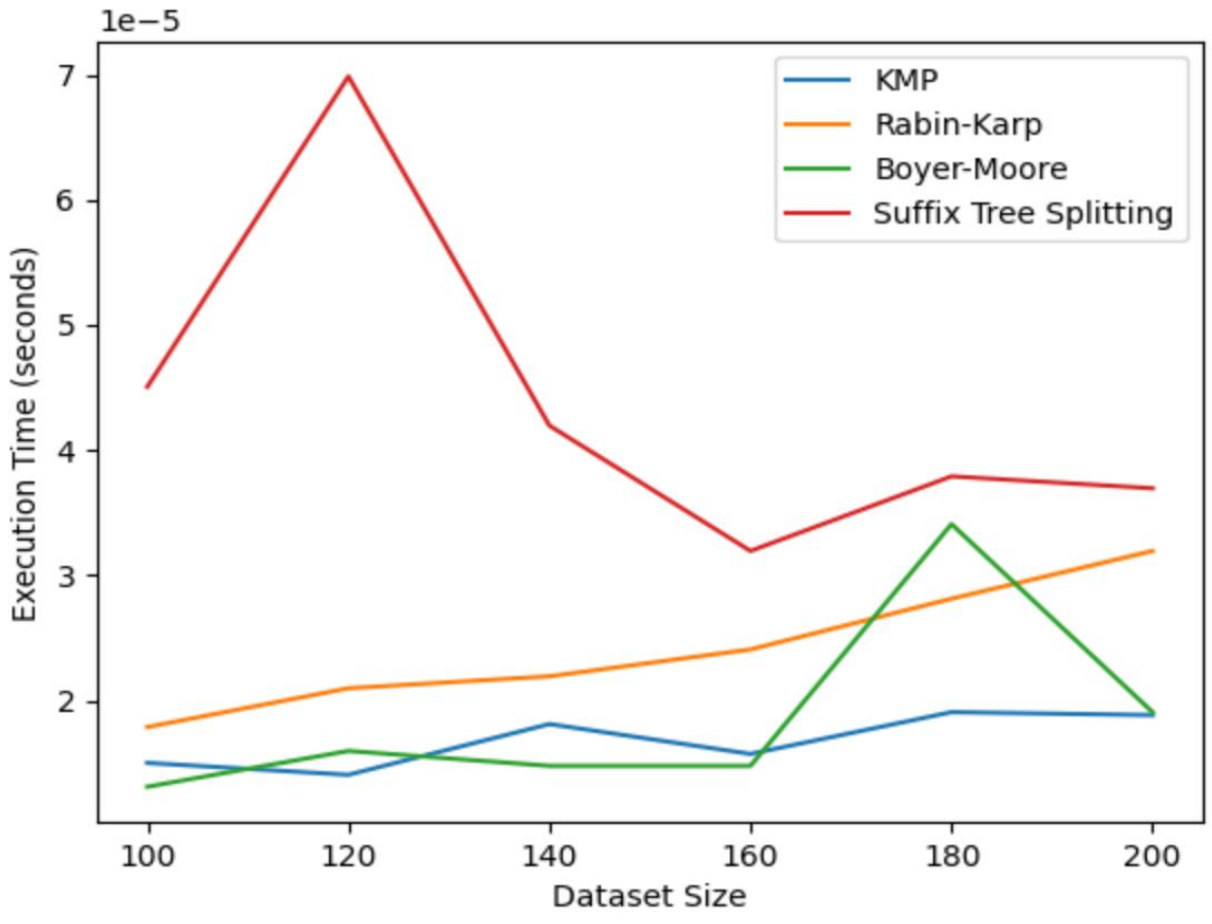

Fig. 8. Data set size (200)

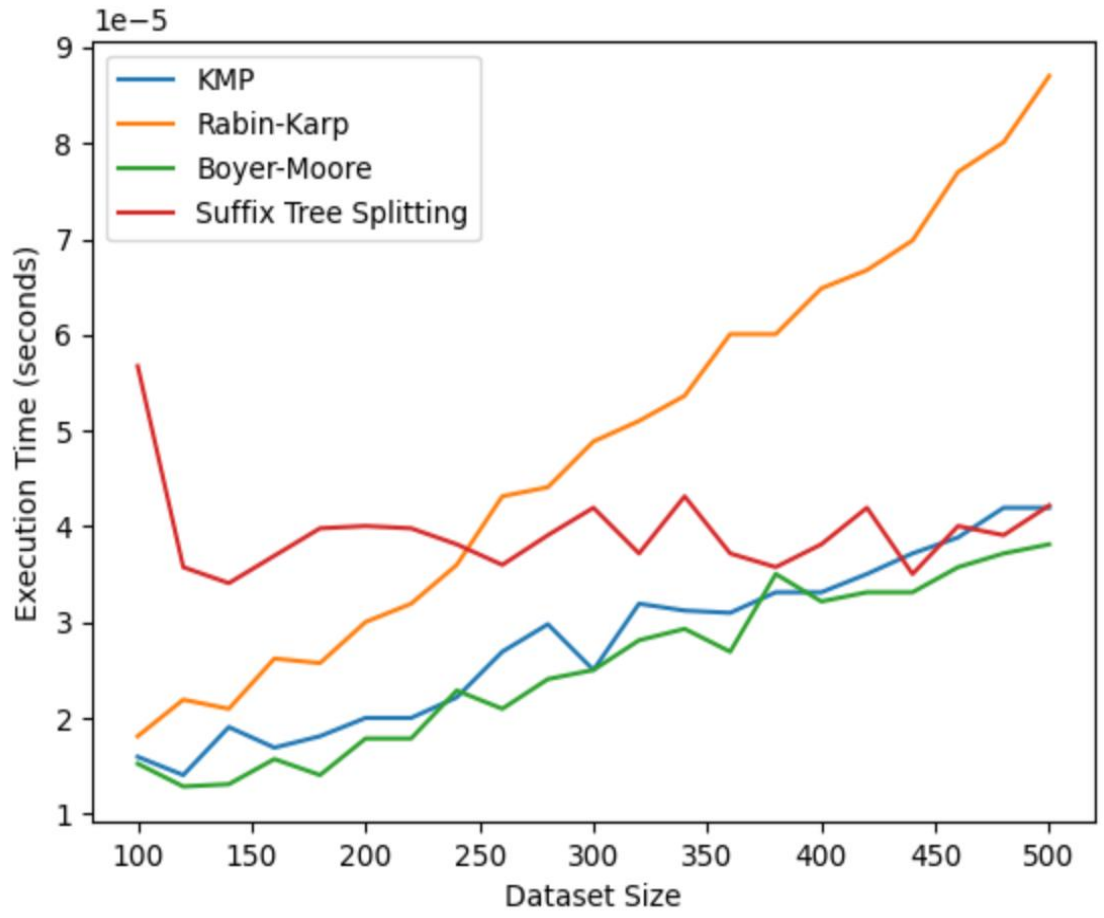

Fig. 9. Data set size (500)

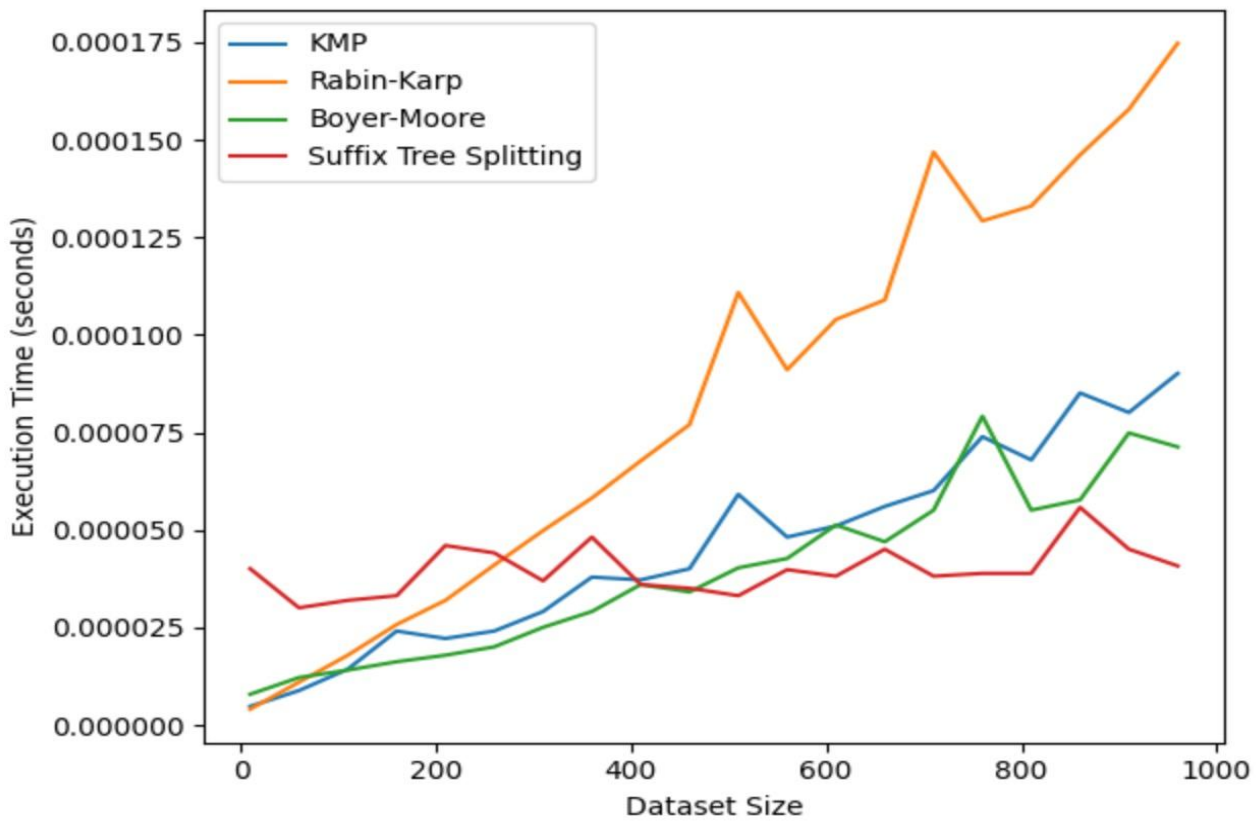

Fig. 10. Data set size (1000)

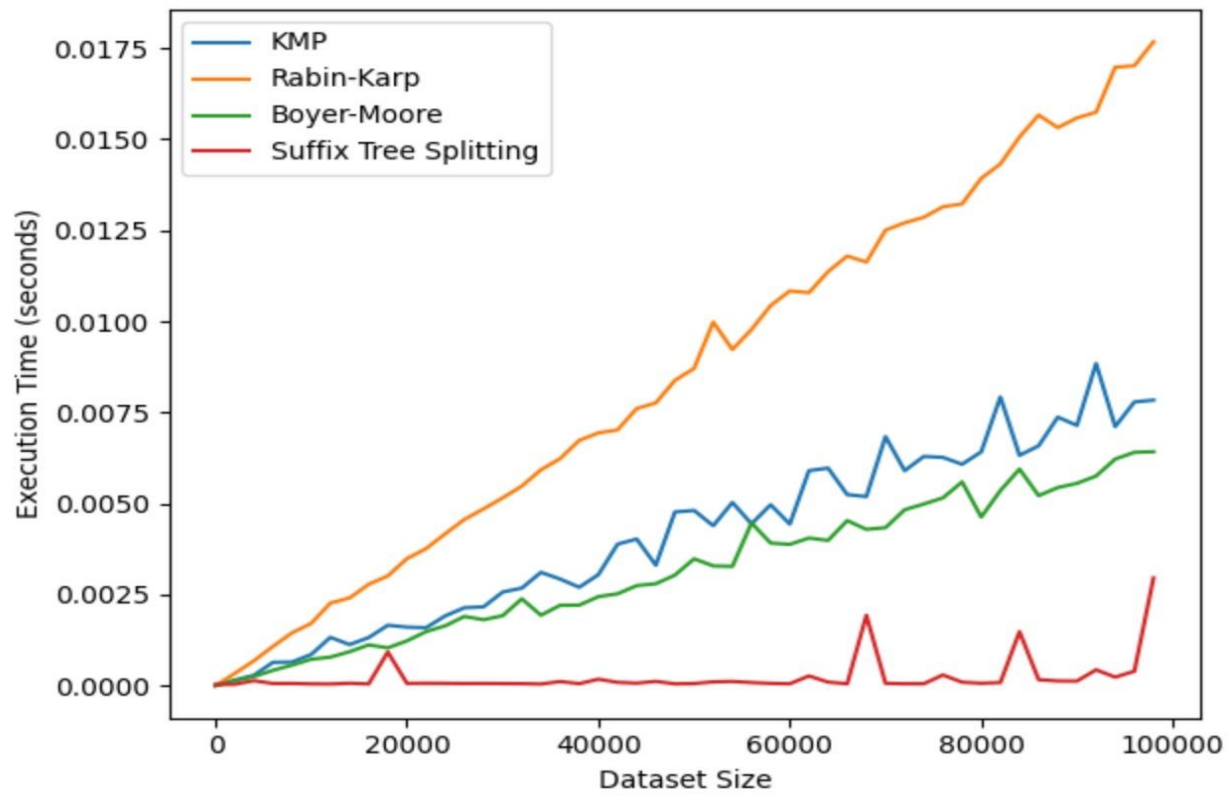

Fig. 11. Data set size (100000)

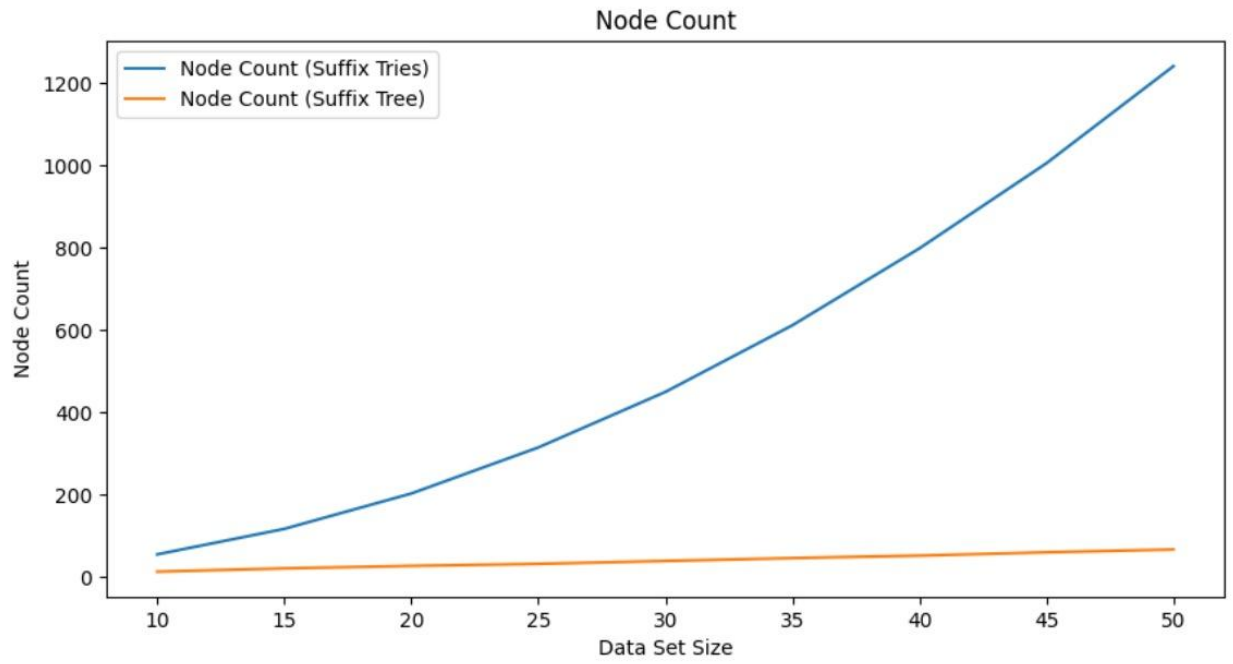

Fig. 12. Node usage

*2) System Resource Utilization*

In the system resource utilization experiment, we fixed the pattern size at 10 characters and analyzed the growing datasets of the Reuters corpus from the NLTK library. We measured tree node count, CPU utilization, and memory usage. A notable finding was the significant difference in node generation between Suffix Tries and Suffix Trees; for instance, the Suffix Tree needed only 19 nodes for 'Mississippi', compared to 66 by Suffix Tries. Regarding memory usage, both algorithms performed similarly for datasets under 500 characters, with no substantial difference due to the 16GB total RAM capacity. The appendices recorded consistent CPU usage across various algorithms, with slight variations in Naive and KMP algorithms.The relevant results are shown in Figure 12 and 13.

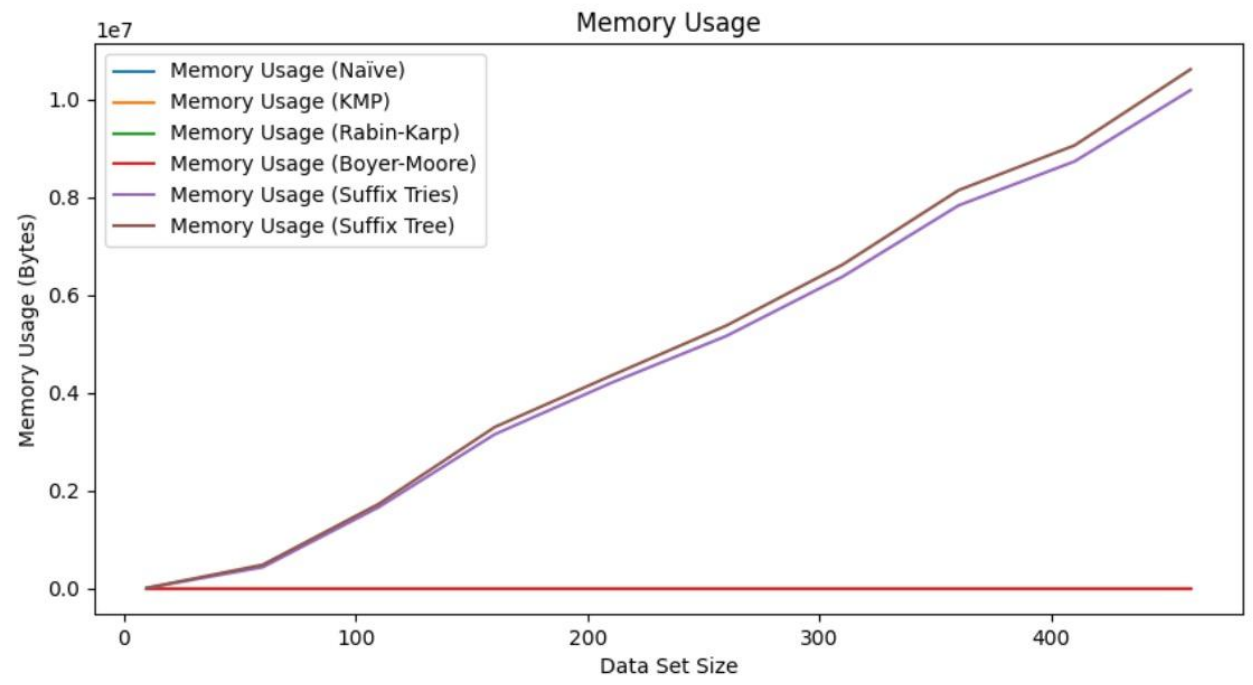

Fig. 13. Memory usage within 500 characters

# V. CONCLUSION

This research delves into text-search algorithms, with a keen focus on optimizing Suffix Trees using Python. Key methodologies explored include the Splitting Method and Ukkonen's Algorithm, with empirical evaluations conducted on diverse datasets like the Reuters corpus and human genomic sequences. The study reveals that while the Splitting Method has higher complexity, Ukkonen's Algorithm offers linear time and space efficiencies. The core finding is the enhanced performance achieved by integrating Ukkonen's Algorithm with a novel search algorithm, creating a synergistic effect that enhances computational efficiency and scalability for text-search challenges. This optimized solution demonstrates superior performance over traditional algorithms such as Naive Search and KMP, especially in terms of resource utilization and algorithmic efficiency. Notably, the research achieves a 100% accuracy rate in pattern detection within genomic sequences, highlighting its potential in bioinformatics. The study underscores the optimized Suffix Tree algorithm as a significant advancement in text-search algorithms, promising valuable applications in fields like natural language processing and database searches. Future research directions include evaluating the new Suffix Tree algorithm against DFS-based methods, exploring Non-deterministic Finite Automata (NDFAs), and enhancing pattern matching in genomic sequences through optimized data structuring strategies.

# REFERENCES

[1] Reyna, J., Hanham, J., & Meier, P. (2018). The Internet explosion, digital media principles and implications to communicate effectively in the digital space. E-Learning and Digital Media, 15(1), 36-52. https://doi.org/10.1177/2042753018754361

[2] S. Akhter and M. A. Haque, "ECG comptression using run length encoding," in 2010 18th European Signal Processing Conference, 23-27 Aug. 2010 2010, pp. 1645-1649.

[3] S. Faro and T. Lecroq, "The exact online string matching problem," ACM Computing Surveys, vol. 45, no. 2, pp. 1–42, 2013.

[4] Nimisha Singla and Deepak Garg. String matching algorithms and their applicability in various applications. International journal of soft computing and engineering, 1(6):218–222, 2012.

[5] Yuebin Bai and H. Kobayashi. New string matching technology for network security. In 17th International Conference on Advanced Information Networking and Applications, 2003. AINA 2003., pages 198–201, 2003.

[6] Donald E. Knuth, James H. Morris, Jr., and Vaughan R. Pratt. Fast pattern matching in strings. SIAM Journal on Computing, 6(2):323–350, 1977.

[7] Robert S. Boyer and J. Strother Moore. A fast string searching algorithm. Commun.ACM, 20(10):762–772, oct 1977.

[8] Zhu Rui Feng and T. Takaoka. On improving the average case of the boyer-moore string matching algorithm. J. Inf. Process., 10(3):173–177, jul 1988.

[9] Ain Zubaidah Mohd Saleh, Nur Amizah Rozali, Alya Geogiana Buja, Kamarularifin Abd. Jalil, Fakariah Hani Mohd Ali, and Teh Faradilla Abdul Rahman. A method for web application vulnerabilities detection by using boyer-moore string matching algorithm. Procedia Computer Science, 72:112–121, 2015. The Third Information Systems International Conference 2015.

[10] Andysah Putera Utama Siahaan. Rabin-karp elaboration in comparing pattern based on hash data. International Journal of Security and its Applications, 12, 04 2018.

[11] Sun Wu, Udi Manber, et al. A fast algorithm for multi-pattern searching. University of Arizona. Department of Computer Science Tucson, AZ, 1994.

[12] Beate Commentz-Walter. A string matching algorithm fast on the average. In International Colloquium on Automata, Languages, and Programming, pages 118–132. Springer, 1979

[13] Ricardo Baeza-Yates and Gaston H Gonnet. A new approach to text searching. Communications of the ACM, 35(10):74–82, 1992.

[14] Esko Ukkonen. On-line construction of suffix trees. Algorithmica, 14(3):249–260,1995.

[15] Gonzalo Navarro, Ricardo A. Baeza-Yates, Erkki Sutinen, and Jorma Tarhio. Indexing methods for approximate string matching. IEEE Data Eng. Bull., 24(4):19–27, 2001.